\documentclass{PoS}

\title{Neutron star long term cooling -\\ Joule heating in magnetized neutron stars}

\ShortTitle{Joule heating governing the cooling of magnetized neutron stars}

\author{\speaker{Deborah N. Aguilera}%
         \thanks{Present address: Theoretical Physics, 
Tandar Laboratory, Comisi\'on Nacional de Energ\'ia At\'omica (CNEA-CONICET), Av. Gral. Paz 1499, 
1650 San Mart\'in, Pcia. Buenos Aires, Argentina}\\
        Departamento de F\'isica Aplicada.  Universidad de Alicante, Spain\\
        E-mail: \email{deborah.aguilera@ua.es}}
\author{Jos\'e A. Pons\\
   Departamento de F\'isica Aplicada.  Universidad de Alicante, Spain\\
       E-mail: \email{jose.pons@ua.es}
}
\author{Juan A. Miralles\\
 Departamento de F\'isica Aplicada.  Universidad de Alicante, Spain\\
        E-mail: \email{ja.miralles@ua.es}
}

\abstract{
          We present two-dimensional simulations for the cooling of neutron stars with strong magnetic fields 
($B \geq 10^{13}$ G).  
We study how the cooling curves are influenced by magnetic field decay. 
We show that the Joule heating effects are very large and in some cases  
control the thermal evolution. 
We characterize the temperature anisotropy induced by the magnetic field and predict the surface
temperature distribution for 
the early and late stages of the evolution of isolated neutron stars, comparing our results 
with available 
observational data of isolated neutron stars.
          }

\FullConference{Supernovae: lights in the darkness\\
		 October 3-5, 2007\\
		 Ma\'o (Menorca) }

\newcommand{\bea}{\begin{eqnarray}}
\newcommand{\eea}{\end{eqnarray}}
\newcommand{\aap}{A\&A}
\newcommand{\apj}{ApJ}
\newcommand{\apjl}{ApJ Lett.}
\newcommand{\apss}{Ap\&SS}
\newcommand{\apjs}{ApJS}
\newcommand{\rxdieciocho}{\hbox{RX~J1856.5$-$3754~}}
\newcommand{\rxcerosiete}{\hbox{RX~J0720.4$-$3125~}}
\newcommand{\rbdoce}{\hbox{RBS~1223~}}

\begin{document}
\section{Introduction}

Recently, observational data of thermally emitting isolated neutron stars (NSs) 
confirm that most of them have magnetic fields larger than $10^{13}$ G. 
Therefore, a reliable treatment of the thermal evolution must
not avoid the inclusion of the effects produced by the presence of 
 high magnetic fields. 
 
The non-uniform distribution of the surface temperature of isolated NSs
seems to be confirmed by the analysis of observational data
(see reviews \cite{Zavlin2007} and \cite{Haberl2007}). 
The mismatch between the extrapolation 
to low energy of the fits to X-ray spectra,
and the observed Rayleigh Jeans tail in the optical band ({\it optical excess flux}),
cannot be addressed with an uniform temperature. 
Several simultaneous fits to multiwavelength spectra of
 \rxdieciocho \cite{Pons2002}, 
\rbdoce \cite{Schwope2007},  
and 
\rxcerosiete \cite{Perez2006} 
are explained by a small hot emitting 
area $\simeq$ 10--20 km$^2$ and an extended cooler component.

It has been proposed that the non-uniform surface temperature distribution may be produced by crustal confined magnetic 
fields \cite{Geppert2004,Azorin2006}. Magnetic fields larger than $10^{13}$~G
 limit the movement of the electrons (heat carriers) in the direction perpendicular to the field 
 with the result that the thermal conductivity is highly suppressed, while remains almost 
 unaffected along the field lines. 
 
 For such large fields, the field decay through Ohmic dissipation and Hall drift processes 
 is very efficient and the heat released in the crust (Joule heating) must be taken into account 
 in the thermal evolution of a neutron star \cite{Aguilera2007}.  In this article we focused on 
 the effects of field decay and Joule heating on the neutron star cooling. 
 In particular, we compare our simulations with observational data of a sample of 
 isolated NSs that are 
 highly magnetized.

\section{Cooling of neutron stars with magnetic fields}

We have performed two--dimensional simulations by solving
the energy balance equation that describes the thermal evolution of a neutron star (NS)

\begin{equation}
C_{v}  \frac{\partial T}{\partial t}  - \vec{\nabla} \cdot
( \hat{\kappa} \cdot \vec{\nabla}  T) = - Q_{\nu} + Q_{\rm J}~,
\label{eneq}
\end{equation}

where $C_v$ is the specific heat per unit volume, $Q_{\nu}$ are energy losses
by $\nu$-emission, $Q_{\rm J}$ the heat released by Joule heating, 
and $\hat \kappa$ is the thermal conductivity tensor, in general anisotropic in presence of
a magnetic field. In this equation we have omitted relativistic factors for simplicity.
A detailed description of the formalism, the code, and results can be found in
\cite{Aguilera2007}. The geometry of the magnetic field is fixed during the evolution.
As a phenomenological description of the field decay, we have assumed the following law 
\begin{equation}
B= B_0\frac{\exp{(-t/\tau_{\rm Ohm})}}
{1+(\tau_{\rm Ohm}/\tau_{\rm Hall})
(1-\exp{(-t/\tau_{\rm Ohm})})}~,
\label{Btime}
\end{equation}
where $B$ is the magnetic field at the pole, $B_0$ its initial value,
 $\tau_{\rm Ohm}$ is the Ohmic characteristic time,
and $\tau_{\rm Hall}$ the typical timescale of the fast, initial Hall stage.
In the early evolution, when $t\ll \tau_{\rm Ohm}$,
we have $B \simeq  B_0 (1+t/\tau_{{\rm Hall}})^{-1}$
while for late stages, when $t \geq \tau_{\rm Ohm}$,
$B \simeq  B_0 \exp(-t/\tau_{{\rm Ohm}})$.
This simple law reproduces qualitatively the results from more complex simulations
\cite{PonsGeppert2007} and facilitates the implementation of field decay
in the cooling of NSs for different Ohmic and Hall timescales.

\section{Joule heating governing the cooling}

Given a $1.35$~M$_{\odot}$ neutron star model with a 
crustal confined magnetic field as in \cite{Aguilera2007}, 
we have varied the parameters
that describe the typical timescales for Ohmic dissipation and a 
fast initial decay induced by the Hall drift.
\begin{figure}[htb]
   \centering
   \vspace{0.9cm}
   \includegraphics[angle=-90,width=0.65\textwidth]{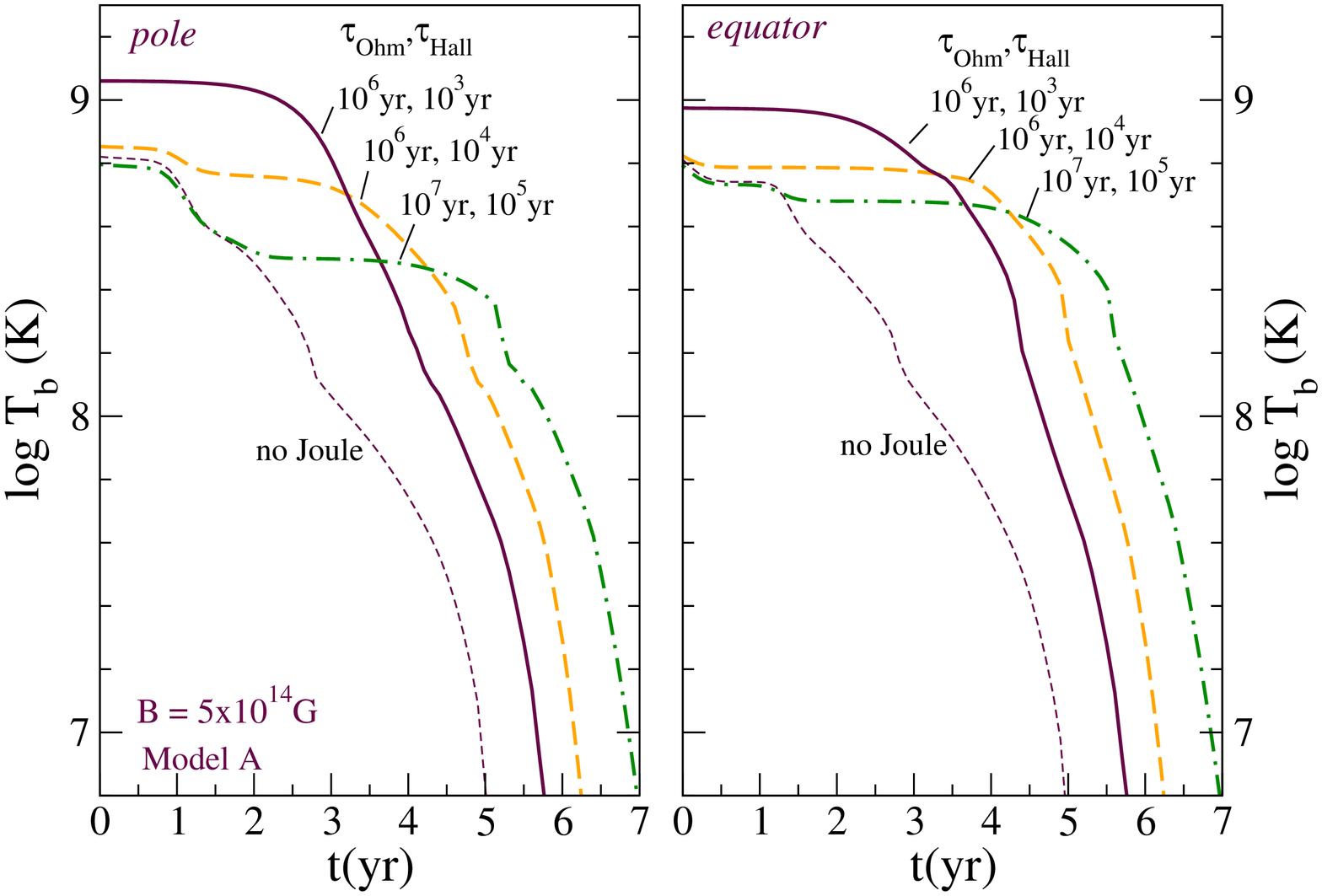}\\
   \vspace{1.3cm}
      \includegraphics[angle=-90,width=0.65\textwidth]{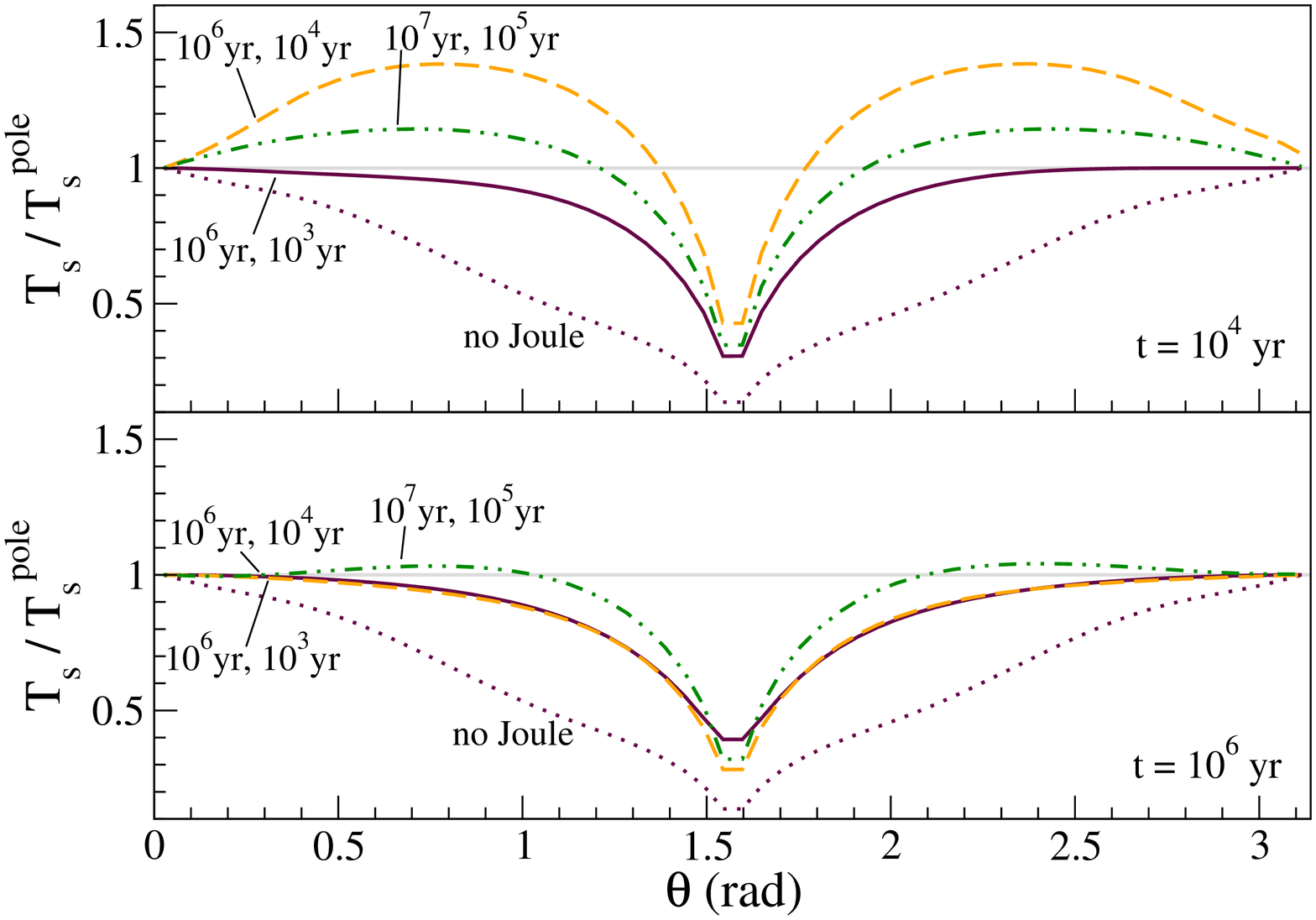}
      \vspace{0.5cm}
\caption{Cooling of strongly magnetized NSs with Joule heating for $B_0=5\times 10^{14}$~G. 
Upper panel shows $T_b$ vs. $t$ at the pole (left) and at the equator (right). 
Lower panel shows
$T_s$ normalized to its value at the pole vs. $\theta$. 
Three pairs of Joule parameters  
($\tau_{\rm Ohm}, \tau_{\rm Hall}$) are shown: 
($10^6$yr,~$10^3$yr) with solid lines, 
($10^6$yr,~$10^4$yr) with dashed lines, and
($10^7$yr,~$10^5$yr) with dotted dashed lines, respectively
}
\label{Joule}
\end{figure}

In Fig.~\ref{Joule} we show the cooling curves for different values of Joule parameters 
($\tau_{\rm Ohm}, \tau_{\rm Hall}$): 
($10^6$yr,~$10^3$yr), 
($10^6$yr,~$10^4$yr), and
($10^7$yr,~$10^5$yr)
represented by
solid lines, dashed lines, and dash-dotted lines, respectively.
For comparison, the thin dashed lines show the evolution with constant field for
the same initial field $B_0=5 \times 10^{14}$ G.

It is first to notice that there is a large effect of the field decay 
on the temperature at the bottom of the envelope $T_b$: as a consequence of the heat released, 
it remains much higher than in the case of non--decaying magnetic field. 
The strong influence of the field decay is evident for all parameters chosen. 
The temperature of the initial plateau is higher for shorter $\tau_{\rm Hall}$,
but the duration of this stage with nearly constant temperature is also shorter. 
When $t=\tau_{\rm Hall}$,   $B$ has decayed to about $1/2B_0$ and $3/4$
of the initial magnetic energy has been dissipated. 
After $t = \tau_{\rm Hall}$,  $T_b$ drops due to the transition 
from the fast Hall decay to the slower Ohmic decay.

The insulating effect of tangential magnetic fields is twofold.  
First, in the absence of additional heating sources, it decouples low latitude regions from the
hotter core resulting in lower temperatures at the base of the envelope. Second; 
if there is heat released in the crust, it prevents the extra heat to flow into
the inner crust or the core where it is more easily lost in the form of neutrinos.
Our simulations with Joule heating show systematically a hot equatorial 
belt at the crust--envelope interface.
However, as discussed in \cite{Aguilera2007}, the
{\it inverted temperature distribution} at the level of the crust
is not necessarily visible in the surface
temperature distribution because it is filtered by the magnetized envelope.
An analysis of the angular temperature distribution given in the 
lower panel of Fig.~\ref{Joule} shows  the development of a middle latitude region 
hotter than the pole at relatively late stages in the evolution ($t \simeq 10^4, 10^5$ yr).
This hotter area is found with a wide range of parameters, and it would have implications
on the light curves of rotating NSs, that will differ substantially from the light curves
obtained with a 
typical hot polar cap model.

\section{Comments on the spin-down age of NSs}
\label{Spindown}
Another aspect that should be reconsidered when we try to fit cooling curves 
to observations is that for many objects the age is calculated
from the measurements of the rotational period $P$ and its 
derivative $\dot P$. The {\em spin down age}, $t_{sd}=P/2\dot P$, is derived
assuming that the lose of angular momentum is entirely due to dipolar radiation 
from a constant (in time) magnetic dipole.
In the case of a decaying magnetic field, $t_{\rm sd}$ can 
seriously overestimate the {\em true age} $t$.
A simple algebra shows that in the case of purely Ohmic decay 
\bea
t= \frac{\tau_{\rm Ohm}}{2} \ln{\left(1+2\frac{t_{\rm sd}}{\tau_{\rm Ohm}}\right)}. 
\eea
In the case that a Hall--induced fast decay also occurs, Eq.~(\ref{Btime}) results in a
large correction of $t_{\rm sd}$ as follows:
\bea
t_{\rm sd}= \tau_{\rm Hall}~f(t)~e^{2t/\tau_{\rm Ohm}} 
\left[f(t) - e^{-t/\tau_{\rm Ohm}}
- \frac{\tau_{\rm Hall}}{\tau_{\rm Ohm}} f(t)\ln{f(t)}\right]
\label{tsd} 
\eea
where 
$f(t)=1+\frac{\tau_{\rm Ohm}}{\tau_{\rm Hall}}(1-e^{-t/\tau_{\rm Ohm}})$. This relation  
gives $t_{\rm sd} \gg t$ by several orders of 
magnitude for  $t \gg \tau_{\rm Hall}$, as shown in Fig.~\ref{Spindown}. 
\begin{figure}[hbt]
   \centering
   \vspace{0.8cm}
   \includegraphics[angle=-90,width=0.75\textwidth]{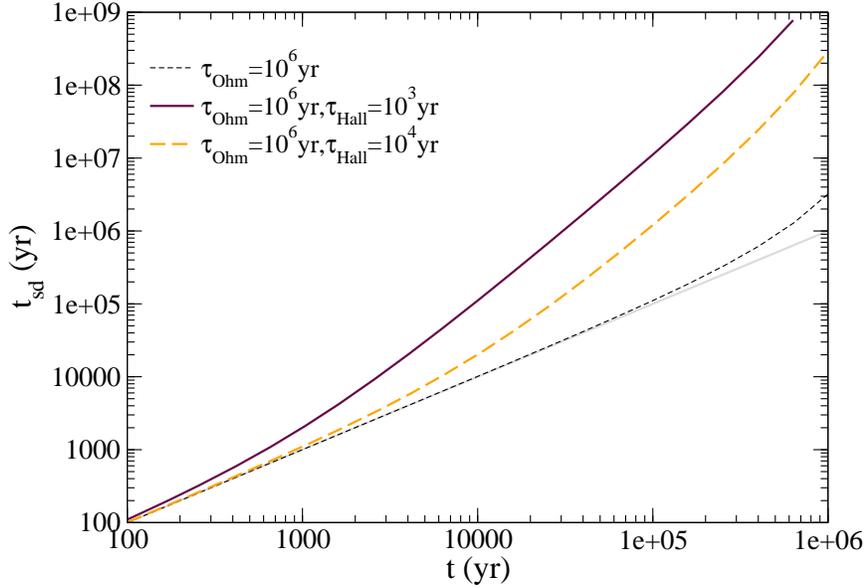}
   \vspace{0.5cm}
\caption{Spin down age ($t_{\rm sd}$) vs. {\em true age} ($t$) 
for a model with magnetic field decay. 
Results for two rates 
$(\tau_{\rm Ohm}, \tau_{\rm Hall})$ are shown: 
($10^6$yr,~$10^3$yr) with solid lines and
($10^6$yr,~$10^4$yr) with dashed lines. 
The grey solid line represents  $t_{\rm sd}=t$, 
and the short dashed line the purely Ohmic decay with $\tau_{\rm Ohm}=10^6$~yr. 
}
\label{Spindown}
\end{figure}
Therefore, the cooling evolution time should be corrected according
to the prescription for the magnetic field decay 
in order to compare with the observations properly. 
A detailed comparison with observational sources is presented in \cite{Aguilera2008} and is 
summarized in the next section. 

\section{Comparison with observations}

\begin{figure}[htb]
   \centering
   \vspace{0.9cm}
   \includegraphics[angle=-90,width=0.9\textwidth]{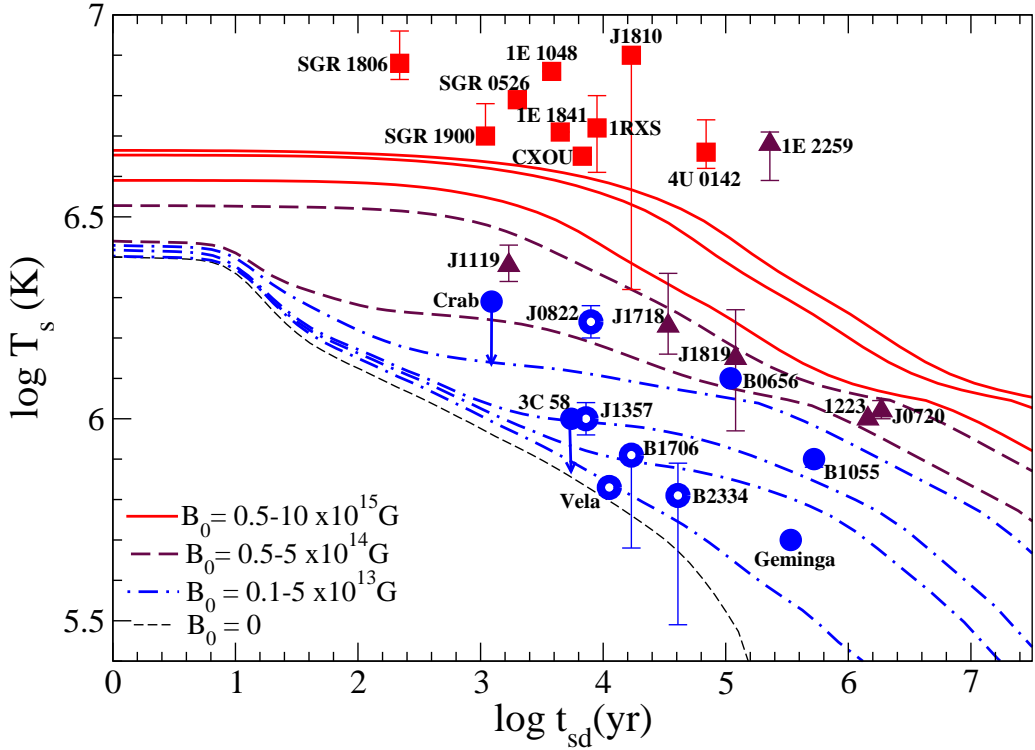}
   \vspace{0.5cm}
\caption{Cooling curves with corrected spin down age. Observational sources correspond
to Table 1 of \cite{Aguilera2008}. Symbols identify sources with the same order of magnetic 
field: squares for magnetars (AXPs and SGRs, with $B \simeq 10^{14-15}$~G), 
triangles for radio-quiet isolated NSs (with $B \simeq 10^{13}$~G), 
and circles for radio-pulsars (with $B \simeq 10^{12}$~G). 
Open circles denote temperatures obtained 
from fits to Hydrogen atmospheres. 
}
\label{compa}
\end{figure}

We compare in Fig.~\ref{compa} our simulations with observational data of NSs covering 
about three orders of magnitude 
in magnetic field strength: from radio-pulsars ($B \simeq 10^{12}$~G) and isolated radio-quiet NSs 
($B \simeq 10^{13}$~G) to recent magnetar candidates ($B \simeq 10^{14-15}$~G). 
The sources considered here are listed in Table 1 of \cite{Aguilera2008}, 
with the corresponding references.

For most NSs, $B$ is estimated by assuming that the lose of angular momentum
is entirely due to dipolar radiation.  The dipolar component is 
$B_d = 3.2 \times 10^{19} ({P \dot{P}})^{1/2}$ G, where $P$ is the
spin period in seconds, and $\dot{P}$ is its time derivative.
In order to work with an homogeneous sample,
we have included in the comparison only those objects for which
$\dot{P}$  is available and the quoted magnetic field is 
$B_d$
and discarded those sources for which $B$ is inferred by other methods\footnote{For a few 
radio-quiet isolated NSs $B$ 
can also be estimated assuming that observed x-ray absorption features are due to proton cyclotron
lines, but this gives the surface field, which is usually larger than the 
external dipolar component}.

The reported temperatures are in most cases blackbody temperatures, except for 
low field radio-pulsars for which 
we take the temperature consistent with
Hydrogen atmospheres following the criteria in \cite{Page2004}. 
Nevertheless, there are some objects in which the
estimate is an upper limit for the thermal component, like the Crab pulsar.
This is also the case for some magnetars, which show 
large variations in the flux in the soft x-ray
band on a timescale of a few years, indicating that the thermal component must be measured during 
quiescence and that the luminosity during their active periods 
is a result of magnetospheric activity.

The age of a NS is subject to a large uncertainty, but it can be estimated by the 
{\it spin-down age} ($t_{sd}=P/2\dot{P}$), provided that the 
birth spin rate far exceeds the present spin rate and $B$ is considered constant\footnote{For 
some cases an independent {\it kinematic age} is available, which  
does not necessarily coincide with $t_{sd}$.}. As shown in Sec.~\ref{Spindown}, 
if one considers magnetic field decay, $t_{sd}$ seriously
overestimates the age $t$ of the simulations. 
Therefore, we transform the cooling curves to plot $T_s$ vs. $t_{sd}$,
to include the temporal variation of the magnetic field. 

In Fig.~\ref{compa} our results show that the effect of high magnetic field 
($B_0\simeq 10^{14-15}$~G, solid lines)
in the cooling
is important from the very beginning of the NS evolution. The temperature reached
is increased up to a factor of 5 in comparison with a non--magnetized 
model and can be kept nearly
constant for about $10^4$ years.
The effect of Joule heating is very significant and may help to explain 
why magnetars are so hot \cite{Kaspi2007}: 
the high temperatures in the early epoch result in higher electrical 
resistivity and in an faster magnetic field dissipation that releases the heat in the crust. 
In this picture, the thermal evolution of radio-quiet, isolated NSs 
could be represented either by NSs born with intermediate fields in the range of 
$B_0=10^{13}$-$10^{14}$~G (dashed lines) or by magnetars in which the field has already 
decayed in a timescale of $\approx 10^{5-6}$ years. For intermediate field strengths, 
the initial effect is not so pronounced but the star
can be kept much hotter than non-magnetized NSs from $10^4$~yr to $10^6$~yrs.
For weakly magnetized NSs, radiopulsars with $B \simeq 10^{12}$~G, the effect of the magnetic
field is small  (dashed dotted lines) and they can satisfactorily be explained 
by non-magnetized models, with the exception of 
 very old objects ($t> 10^6$ years), as discussed in \cite{Miralles98}.

\section{Conclusions}
From the results presented here we conclude that the thermal evolution of a magnetized NSs is 
strongly affected by the presence of magnetic fields. Therefore, studies aimed to disentangle 
properties of NSs interior (e.g. EoS, neutrino processes, etc.) 
for objects
with $B \geq 10^{13}$, 
through cooling curves should not neglect the (dominant) magnetic field effects. 
A first step towards a coupled magneto-thermal evolution has been given in this work and 
future investigations will consider a consistent evolution including the evolution of the magnetic 
field geometry.

\subsubsection{Acknowledgements}

D.N.A was supported by the VESF fellowship EGO-DIR-112/2005.
This work has been supported in part by the Spanish MEC grant AYA 2004-08067-C03-02.


\begin{thebibliography}{99}

\bibitem{Zavlin2007}
{Zavlin}, V.~E. 2007, preprint [astro-ph/0702426]

\bibitem{Haberl2007}
{Haberl}, F. 2007, \apss, 308, 181

\bibitem{Pons2002}
{Pons}, J.~A et al. 2002, \apj, 564, 981 

\bibitem{Schwope2007}
{Schwope}, A.~D. and {Hambaryan}, V. and {Haberl}, F. and {Motch}, C. 2007, \apss, 308, 619


\bibitem{Perez2006}
{P{\'e}rez-Azor{\'{\i}}n}, J.~F., {Pons}, J.~A., {Miralles}, J.~A., \&
  {Miniutti}, G. 2006, \aap, 459, 175 

\bibitem{Geppert2004}
{Geppert}, U., {K{\"u}ker}, M., \& {Page}, D. 2004, \aap, 426, 267

\bibitem{Azorin2006}
{P{\'e}rez-Azor{\'{\i}}n}, J.~F., {Miralles}, J.~A., \& {Pons}, J.~A.
  2006, \aap, 451, 1009

\bibitem{Aguilera2007}
{Aguilera}, D.~N., {Pons}, J.~A., \& {Miralles}, J.~A., \aap~2008 (in press),  
preprint [0710.0854] (astro-ph).

\bibitem{PonsGeppert2007}
{Pons}, J.~A. \& {Geppert}, U. 2007, \aap, 470, 303

\bibitem{Aguilera2008} Aguilera, D.~N., Pons, 
J.~A., \& Miralles, J.~A.\ 2008, \apjl, 673, L167 

\bibitem{Page2004}
{Page}, D., {Lattimer}, J.~M., {Prakash}, M., \& {Steiner}, A.~W. 2004, \apjs,
  155, 623
  
\bibitem{Kaspi2007}
{Kaspi}, V.~M. 2007, \apss, 308, 1

\bibitem{Miralles98}
{Miralles}, J.~A., {Urpin}, V., \& {Konenkov}, D. 1998, \apj, 503, 368  

\end{thebibliography}
\end{document}